\documentclass[a4paper]{article}

\usepackage{INTERSPEECH2015}

\usepackage{graphicx}
\usepackage{amssymb,amsmath,bm}
\usepackage{textcomp}

\newcommand{\Fig}[1]{Fig. \ref{fig:#1}} 
\newcommand{\Figure}[3]{\vspace{-0mm} \includegraphics[width=#1,clip]{#2.eps} \vspace{-0mm} \caption{#3} \vspace{-0mm} \label{fig:#2}}

\newcommand{\Norm}[1]{ {\mathcal N}\left( #1 \right) } 
 
\renewcommand{\Vec}[1]{\textrm{\boldmath $#1$}} 

\newcommand{\Section}[1]{\vspace{-0mm}\section{#1}\vspace{-0mm}}

\newcommand{\pt}[1]{\left(#1\right)} 
\newcommand{\br}[1]{\left[#1\right]} 
\newcommand{\x}{ \Vec{x} } 
\newcommand{\y}{ \Vec{y} } 
\newcommand{\haty}{ \Vec{\hat y} } 
  
\newcommand{\drawfig}[4]{ 
  \begin{figure}[#1]
  \begin{center}
  \Figure{#2}{#3}{#4} 
  \end{center} 
  \end{figure}
}

\ninept

\title{Sampling-based speech parameter generation\\using moment-matching networks}

\makeatletter
\def\name#1{\gdef\@name{#1\\}}
\makeatother
\name{{\em Shinnosuke Takamichi$^{1}$, Tomoki Koriyama$^{2}$, Hiroshi Saruwatari$^{1}$}}

\address{$^1$ Graduate School of Information Science and Technology, The University of Tokyo, Japan. \\
$^2$ Interdisciplinary Graduate School of Science and Engineering, Tokyo Institute of Technology, Japan. \\
  {\small \tt \{shinnosuke\_takamichi,hiroshi\_saruwatari\}@ipc.i.u-tokyo.ac.jp, koriyama@ip.titech.ac.jp}
}

\begin{document}

\maketitle

\begin{abstract}
  This paper presents sampling-based speech parameter generation using moment-matching networks for Deep Neural Network (DNN)-based speech synthesis. Although people never produce exactly the same speech even if we try to express the same linguistic and para-linguistic information, typical statistical speech synthesis produces completely the same speech, i.e., there is no inter-utterance variation in synthetic speech. To give synthetic speech natural inter-utterance variation, this paper builds DNN acoustic models that make it possible to randomly sample speech parameters. The DNNs are trained so that they make the moments of generated speech parameters close to those of natural speech parameters. Since the variation of speech parameters is compressed into a low-dimensional simple prior noise vector, our algorithm has lower computation cost than direct sampling of speech parameters. As the first step towards generating synthetic speech that has natural inter-utterance variation, this paper investigates whether or not the proposed sampling-based generation deteriorates synthetic speech quality. In evaluation, we compare speech quality of conventional maximum likelihood-based generation and proposed sampling-based generation. The result demonstrates the proposed generation causes no degradation in speech quality.
\end{abstract}

\Section{Introduction}
    Statistical speech synthesis \cite{zen09} is a method of synthesizing speech using statistical models. One of the final goals of speech synthesis is to synthesize speech as {\it natural} as humans' speech. Speech quality is an important factor of naturalness, and various methods have been developed for improving the quality of synthesized speech \cite{takamichi14_RCM,takamichi16mspf,saito17asv}. In particular, Deep Neural Network (DNN)-based speech synthesis \cite{zen13dnn,oord16wavenet} has remarkably improves the quality of synthesized speech. However, speech quality is just one factor of naturalness, so the {\it naturalness} of synthetic speech needs to be evaluated from other perspectives.

  This paper focuses on a new metric for naturalness: inter-utterance variation within a given context \cite{inukai13variation}. As shown in \Fig{eps/nat_vs_syn}, conventional DNN-based speech synthesis produces speech on the basis of a minimum error criterion, and generates the same synthetic speech from the given context. Therefore, when the context is fixed, the synthesized speech is always the same and the repeatedly synthesized speech is just recorded and played back. On the other hand, people never produce exactly the same speech even if we try to fix linguistic and para-linguistic information. We call this variation inter-utterance variation (or intra-speaker variation) \cite{inukai13variation}, and this paper addresses ways to synthesize speech that has inter-utterance variation. The straightforward way is to explicitly model the inter-utterance variation using repeatedly spoken speech data that have the same context. However, typical training data for speech synthesis do not include such data, and collecting the data is time-consuming work. Another way is to randomly sample speech from the conditional probability distribution of speech parameters. Shannon et al. \cite{shannon11normalized} evaluated random sampling from the trajectory Hidden Markov Models (HMMs) \cite{zen07}. They reported that speech quality in randomly-sampled synthetic speech is significantly degraded below that in Maximum Likelihood (ML)-based synthetic speech. One reason for quality degradation is time quantization by HMMs \cite{watts16explorednn}, and sampling from trajectory DNN \cite{hashimoto15} or mixture density networks \cite{zen14} is the better solution. However, in practice, sampling from such distributions is computationally inefficient since a speech parameter sequence is a high dimensional feature. 

        \drawfig{t}{0.98\linewidth}{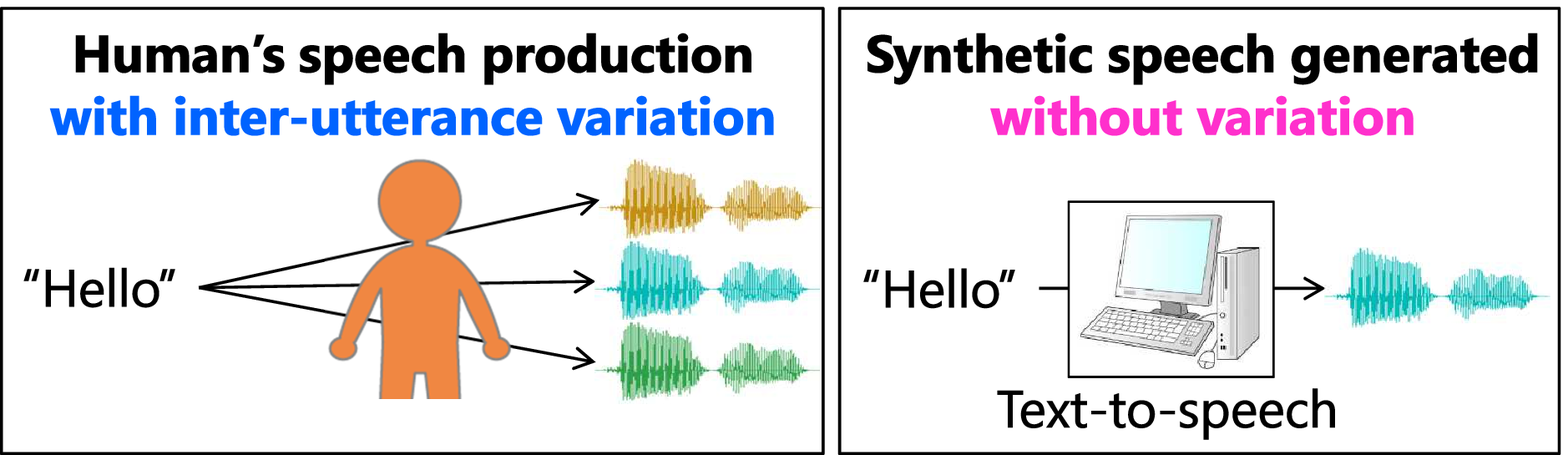}{Comparison of natural and synthetic speech. When input context is fixed, humans' speech has variation between utterances, but conventional synthetic speech does not.}

  This paper proposes sampling-based speech parameter generation using moment-matching networks. DNNs are trained to match moments of natural speech parameters and synthetic speech parameters. The variation of speech parameters is compressed into a low-dimensional simple prior noise vector, and the DNNs transform the simple noise into the speech parameter variation. In synthesis, the synthetic speech parameters are randomly sampled using the sampled prior noise. Since the variation is represented as simple noise, the proposed sampling-based generation is more computationally efficient than the direct sampling of speech parameters. As the first step towards generating synthetic speech that has natural inter-utterance variation, this paper investigates whether or not the proposed sampling-based generation deteriorates synthetic speech quality. In evaluation, we compared speech quality of ML-based generation and proposed sampling-based generation, and the results demonstrate that the proposed generation causes no degradation in speech quality.

\Section{Conventional DNN-based speech synthesis and sampling from acoustic models}
	In conventional DNN-based speech synthesis, the DNN parameters (e.g., matrices and biases) are estimated to minimize the mean squared error between natural speech parameters and generated speech parameters \cite{zen13dnn,wu15dnnmge,zen15}. The training criterion is equivalent to ML-based training with an isotropic Gaussian distribution (Gaussian distribution having an isotropic covariance matrix) and estimates only mean vectors of the Gaussian distribution. In synthesis, the same as in training, the acoustic models generate ML estimates of speech parameters given the input context. Therefore, when the input context is fixed, the generated speech parameters are always the same.
    
    We can randomly sample speech parameters from more proper distributions such as trajectory models \cite{shannon11normalized,hashimoto15}. Although the trajectory models make it possible to sample speech parameters considering static and delta constraints, the computation cost for sampling is too high because the trajectory model has large full-covariance matrices. Also, mixture density networks \cite{zen14} can represent proper distributions such as Gaussian mixture models but such complicated models suffer from high computation cost and low accuracy of sampling.
   
\Section{Sampling-based speech parameter generation using moment-matching networks}
	This section describes how to randomly sample speech parameters from moment-matching networks.
	The training criterion does not assume an explicit distribution form (such as Gaussian \cite{hashimoto15} and mixture models \cite{zen14}) but minimizes the distance between moments of training and generated data.

	\subsection{Moment-matching networks}
      \subsubsection{Minimizing Maximum Mean Discrepancy (MMD) \cite{li15momentmatchingnetwork}}
          Let $\y = \br{\y_1^\top, \cdots, \y_t^\top, \cdots, \y_T^\top}^\top$ and $\haty = \br{\haty_1^\top, \cdots, \haty_t^\top, \cdots, \haty_T^\top}^\top$ be a feature sequence of training data and
          that generated through DNNs. $\y_t$ and $\haty_t$ are training and generated data at frame $t$.
          $T$ is the sequence length. 
          The DNNs that can sample $\haty$ are trained to minimize squared error between
          moments of $\y$ and $\haty$. The criterion is known as the square of (kernelized)
          Maximum Mean Discrepancy (MMD) as follows:
              \begin{align}	
                  L_{\rm MMD} \pt{\y, \haty}
                              = & \frac{1}{T^2} \left\{
                                {\rm tr}\pt{  \Vec{1}_{T} \cdot \Vec{K}_{y}\pt{\y,\y}} \right. \nonumber \\
                                & \hspace{-1.3cm} \left.
                              + {\rm tr}\pt{\Vec{1}_{T} \cdot \Vec{K}_{y}\pt{\haty,\haty}} 
                              - 2 \cdot {\rm tr}\pt{\Vec{1}_{T} \cdot \Vec{K}_{y}\pt{\y,\haty}} \right\},
              \end{align}	   
          where 
          $\rm{tr}\pt{\cdot}$ denotes matrix trace，
          $\Vec{1}_{T}$ is $T$-by-$T$ matrix all components of which are 1.
          $\Vec{K}_{y}\pt{\y,\haty}$ is a gram matrix between $\y$ and $\haty$.
          $\left\{t,\tau\right\}$-th component of $\Vec{K}_{y}\pt{\y,\haty}$ is an arbitrary kernel function between $\y_t$ and $\haty_\tau$.
          If the Gaussian kernel is chosen, 0th-through-infinite order moments are considered in training. 
          The criterion $L_{\rm MMD}$ is used to train DNNs that have a  
          low-dimensional noise vector $\Vec{n}$ as input and 
          $\haty$ as output.
          The DNNs can be regarded as models to transform statistics of $\Vec{n}$ into 
          those of $\y$ as shown in \Fig{eps/matching}.

        \drawfig{t}{0.98\linewidth}{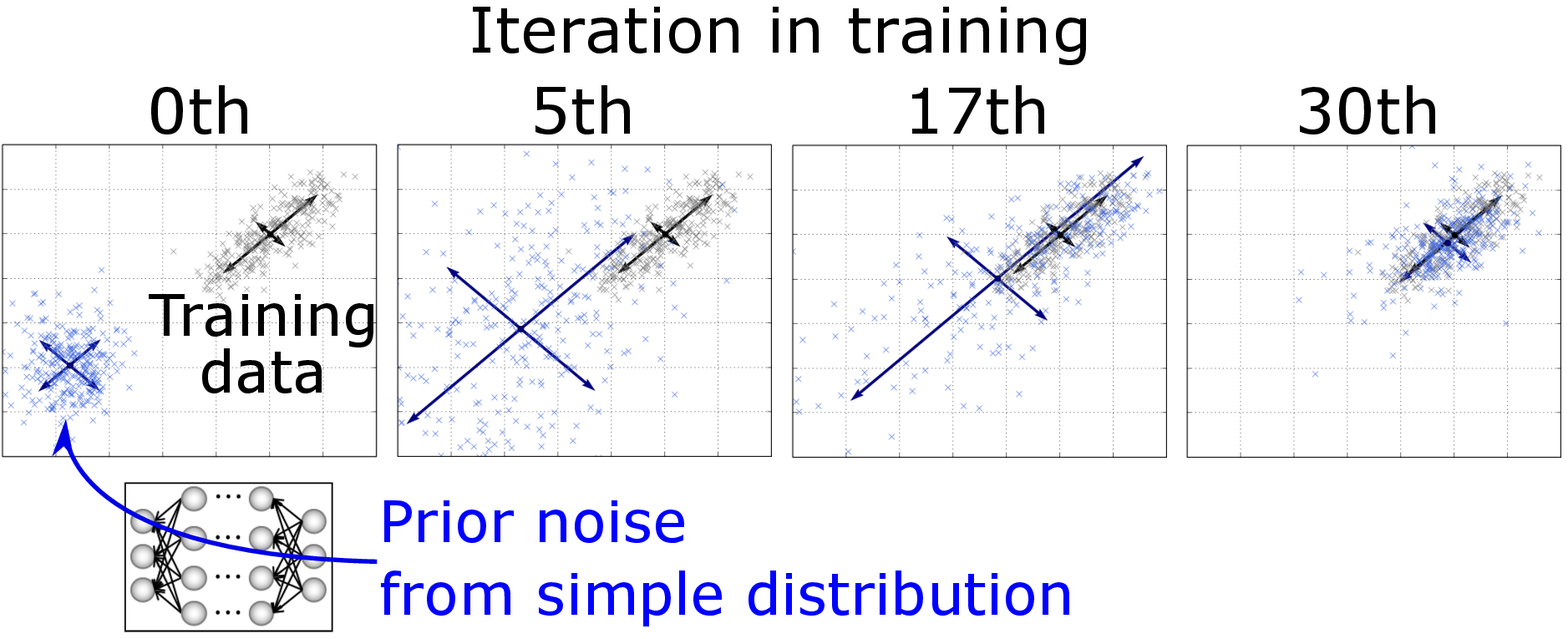}{Training of moment-matching networks. Distributions of training (black) and generated (blue) data are drawn with their 1st and 2nd moments.
        	Networks are trained to transform prior simple noise into data distribution.}

      \subsubsection{Minimizing conditional MMD \cite{ren16conditionalmomentmatching}}
          The method in Section 3.1.1 can be extended to conditional distributions.
          Given the referred to input sequence $\x = \br{\x_1^\top, \cdots, \x_t^\top, \cdots, \x_T^\top}^\top$, the DNNs are trained to match moments of $\y$ and $\haty$. 
          The input of DNNs is a concatenated vector $\tilde{\Vec{x}} = [\x^\top, \Vec{n}^\top]^\top$, 
          and the following loss function called conditional MMD is used to train DNNs,

              \begin{align}	
                  L_{\rm CMMD}\pt{\tilde{\Vec{x}}, \y, \haty}
                              = & \frac{1}{T^2} \left\{
                                {\rm tr}\pt{\Vec{G}\pt{\tilde{\Vec{x}}} \cdot \Vec{K}_y\pt{\y,\y}} \right. \nonumber \\
                                & 
                              + {\rm tr}\pt{\Vec{G}\pt{\tilde{\Vec{x}}} \cdot \Vec{K}_y\pt{\haty,\haty}} \nonumber \\
                                & \left.
                              - 2 \cdot {\rm tr}\pt{\Vec{G}\pt{\tilde{\Vec{x}}} \cdot \Vec{K}_y\pt{\y,\haty}} \right\}, \\
                  \Vec{G}\pt{\tilde{\Vec{x}}} =  & 
                              \tilde{\Vec{K}}_{x}^{-1}\pt{\tilde{\Vec{x}}}
                              \Vec{K}_{x}\pt{\tilde{\Vec{x}}}
                              \tilde{\Vec{K}}_{x}^{-1}\pt{\tilde{\Vec{x}}}, \\
                  \tilde{\Vec{K}}_{x}\pt{\tilde{\Vec{x}}} = &
                              \Vec{K}_{x}\pt{\tilde{\Vec{x}}} + \lambda \Vec{I}_{T},
              \end{align}	 
          where $\Vec{I}_{T}$ is the $T$-by-$T$ identical matrix,
          $\lambda$ is the regularization coefficient, and 
          $\Vec{K}_{x}\pt{\tilde{\Vec{x}}}$ is a gram matrix of $\tilde{\Vec{x}}$.

          In generation, $\haty$ is randomly sampled from DNNs given the referred input $\x$ and randomly sampled $\Vec{n}$. 

	\subsection{Applying moment-matching networks to sampling-based speech parameter generation}
    	Here, we describe how to perform sampling-based speech parameter generation using moment-matching networks.
    	As shown in \Fig{eps/moment_matching}, DNNs are trained to minimize conditional MMD.
        $\x$ and $\y$ are the contextual feature sequence of input text and parameter sequence of output speech, respectively.
        $\Vec{n}$ is sampled frame-by-frame from the simple distribution.
        The DNNs predict static and dynamic features of output speech parameters, 
        and $\haty$ is determined considering these features \cite{tokuda00}.
        In synthesis, after linguistic features are determined and the prior noise is sampled, 
        the generation process is performed in the standard manner \cite{zen13dnn}.
        
        Since most of contextual factors are $1$-of-$K$ hot vectors and their combination becomes an enormous number,
        the kernel function between contextual factors does not have much information.
    	Hence, we used bottleneck features instead of contextual features for kernel calculation.
        Other Feed-Forward networks that predict $\y$ from $\x$ are constructed using a mean squared error criterion \cite{zen13dnn},
        and continuous values of the specific hidden layer are used to calculate kernels.
        
        \drawfig{t}{0.95\linewidth}{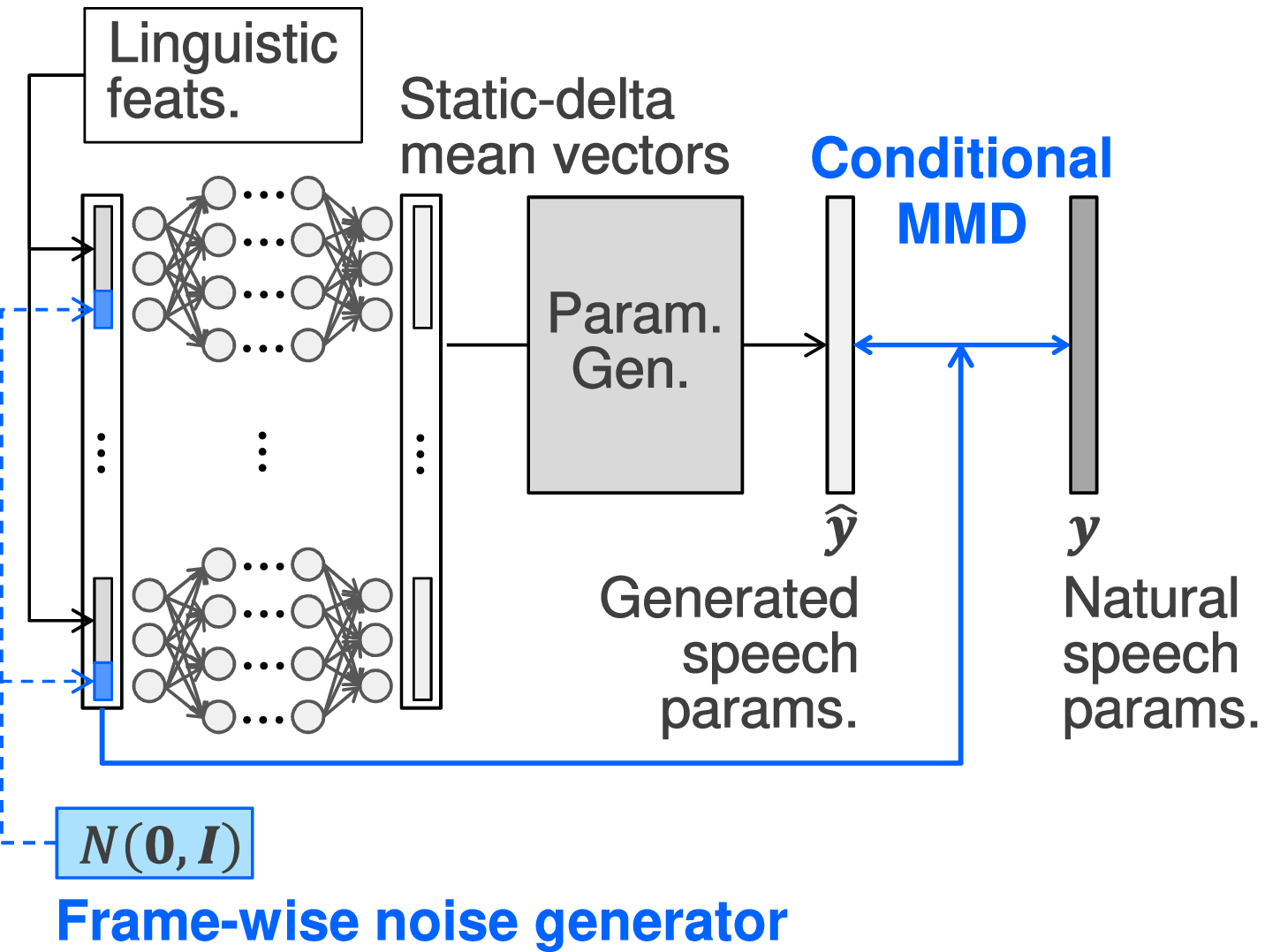}{Sampling-based speech parameter generation using moment-matching networks. Note that linguistic features are directly used in this figure for clear illustration, but bottleneck features \cite{wu15multi} are used in place of linguistic features in actual implementation.}
        
	\subsection{Discussion}
    	Since conditional MMD is a non-parametric measure, our algorithm can model a more complicated distribution than mixture density networks \cite{zen14} or trajectory DNNs \cite{hashimoto15}.
     	Additionally, since the variation of speech parameters is compressed into the low-dimensional prior noise vector, our sampling-based generation is more computationally efficient than conventional models \cite{zen14,hashimoto15}.
        
        A Generative Adversarial Network (GAN) \cite{goodfellow14gan} and conditional GAN \cite{mirza15conditionalgan} are also non-parametric modeling of probability distributions.
        The GAN requires many tricky techniques to optimize since the optimization is a minimax problem \cite{goodfellow16gantutorial}.
        We have already developed speech synthesis integrating GAN \cite{saito17asv}, but the algorithms presented in this paper is 
        easier to optimize than the GAN-based methods. This is because our algorithm is just a minimization problem of conditional MMD.
        Here, we further discuss the relationships among GAN, our algorithm (moment matching), and conventional speech-related techniques.
        GAN minimizes divergence between natural and generation distribution, such as Jensen-Shannon divergence \cite{goodfellow14gan} and $f$-divergence \cite{nowozin16fgan},
        therefore, GAN-based methods are related to divergence-based speech processing such as $\beta$-divergence \cite{fevotte09betadivergence, gustav16betadivergence}.
        On the other hand, moment matching explicitly uses the distance between moments, so our method is 
        related to conventional speech processing methods using higher-order statistics, such as kurtosis \cite{miyazaki12musicalnoise}, global variance \cite{toda07_MLVC} and modulation spectrum \cite{takamichi16mspf}.
        
        A conventional approach to add inter-utterance variation is to use of sentence-level context \cite{watts15sentence}. Whereas this method corresponds to adding speech expressions intentionally added by the speaker, our method adds speech randomness that is not intended by the speaker.
        
        Finally, we discuss whether our algorithm can deceive Anti-Spoofing Verification (ASV) \cite{wu15asvspoof} which detects voice spoofing attacks. Deceiving the ASV has become the metric to evaluate speech synthesis \cite{saito17asv}, and one ASV technique is replay-attack detection \cite{lindberg99replayattack} which determines whether the presented voice is raw or recorded.
        The technique detects recorded voices on the basis of degree of coincidence between the pre-recorded and presented voices.
        Conventional speech synthesis is easily detected by the replay-attack detection because repeatedly synthesized speech is completely the same as the recorded voice.
        On the other hand, we expect that our algorithm can mitigate the detection rate because it has inter-utterance variation as shown in \Fig{eps/mcep_sequence}.
        
        \drawfig{t}{0.98\linewidth}{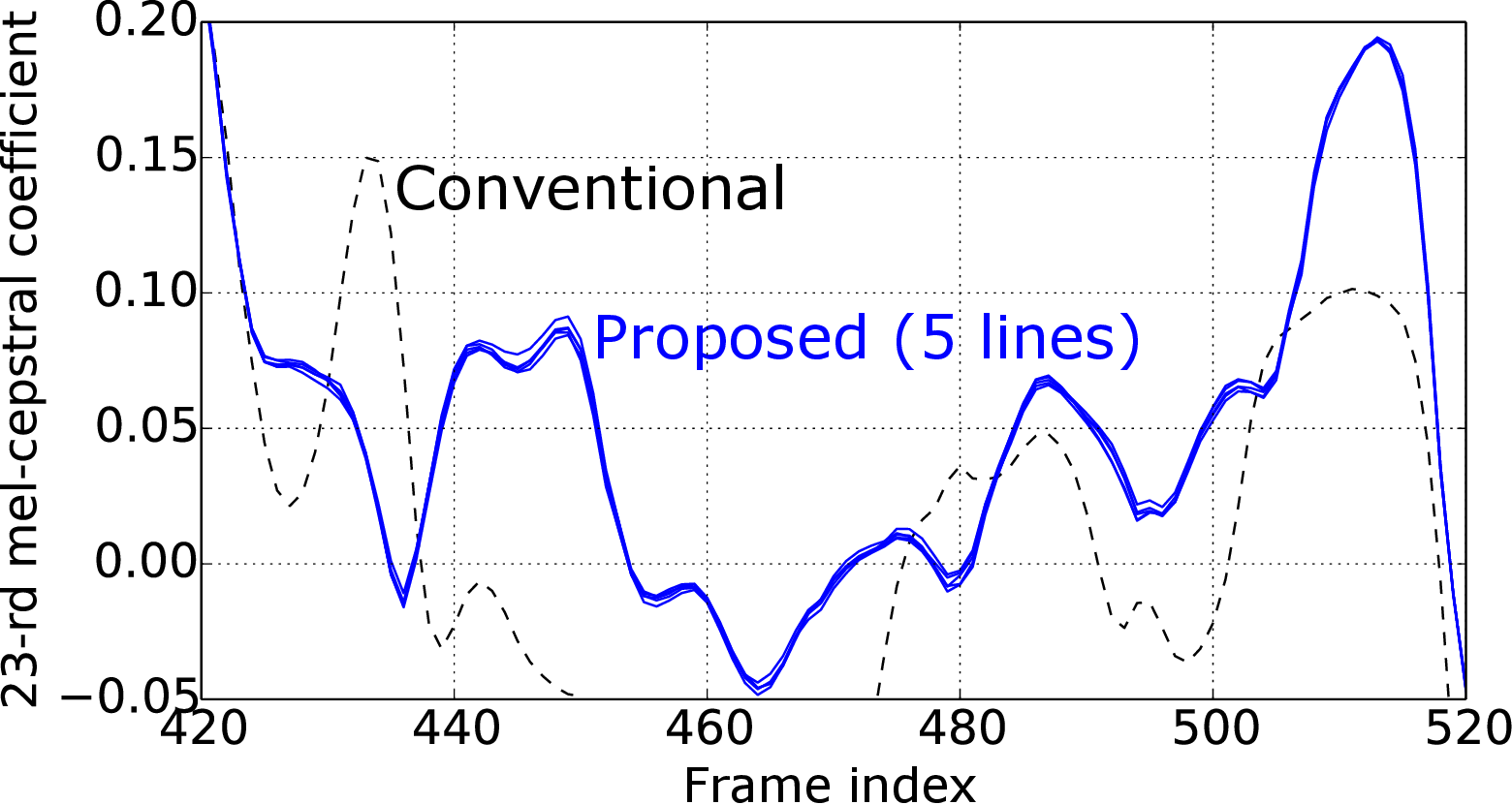}{An example of generated speech parameter trajectories. We sampled five trajectories using proposed method.
        	Conventional one is trained using mean squared error.}

\Section{Experimental evaluation}
	\subsection{Experimental conditions}
      We used speech data of five female speakers taken from the ATR Japanese speech database \cite{abe90}. 
      Each speaker uttered 503 phonetically balance sentences. 
      We used 450 sentences per speaker (2250 sentences in total) for the training
      and 53 sentences of one speaker for the evaluation. Speech signals were
      sampled at a rate of 16 kHz, and the shift length was set to 5 ms. The
      0th-through-24th mel-cepstral coefficients were used as a spectral
      parameter and $F_0$ and 5 band-aperiodicity \cite{kawahara01,ohtani06} were used as excitation
      parameters. The WORLD analysis-synthesis system \cite{kawahara99}
      was used to extract the parameters and synthesize the waveform.
      To improve training accuracy, speech parameter trajectory smoothing \cite{takamichi15blizzard} with a 50 Hz cutoff modulation frequency was applied to the spectral parameters in the training data. 
      Contextual features includes 274-dimensional linguistic features\footnote{We only used some of the prosody-related features because we predicted only the spectral parameters in this experiment.} (phonemes, mora
      position, and so on.) and 5-dimensional speaker codes \cite{hojo16speakercode}.
      The contextual features and speech features are normalized to have zero-mean unit-variance,
      and 80\% of the silence frames were removed from the training data.
      The DNN architectures was Feed-Forward networks that includes 3 $\times$ 512-unit Rectified Linear
      Unit (ReLU) \cite{glorot11relu} hidden layers and a 75-unit linear output layer.
      The DNN predicted only spectral parameters, and $F_0$ and band aperiodicity of 
      natural speech were used for waveform synthesis.
      The input of DNNs was 128-dimensional bottleneck features and a 3-dimensional noise vector per frame.
      The noise is sampled from $\Norm{\Vec{0}, \Vec{I}_3}$
      The regularization coefficient $\lambda$ was set to $0.01$.
      A Gaussian kernel was used as a kernel function for speech parameters $\y$, i.e., ${\rm exp}\{-\|\y_t - \haty_\tau\|^2 / \sigma^2 \}$.
      $\sigma$ was set so that $\|\y_t - \haty_\tau\|^2 < 1$ for all the training data \cite{ren16conditionalmomentmatching}.
      The same kernel settings were used for the input features.
      
      As described in Section 1, investigating naturalness of inter-utterance variation is our next work, and
      this paper investigates the effect of sampling-based generation in speech quality \cite{shannon11normalized}.
      We evaluated the following speech synthesis systems:
      	\begin{description} 
			\item {\bf conv:} conventional speech synthesis minimizing mean squared error \cite{zen13dnn}
            \item {\bf pro (w/ rand):} proposed speech synthesis with sampling-based generation
            \item {\bf pro (w/o rand):} proposed speech synthesis with ML-based generation
		\end{description}
      For ``pro (w/o rand),'' we fixed the prior noise vector to $\Vec{0}$ (ML estimates) in synthesis.
      ``pro (w/o rand)'' is trained the same as ``pro (w/o rand),'' but the speech parameters have no inter-utterance variation.
      
      A preference AB test was conducted by eight listeners. We presented every pair of generated speech of three systems in random order, and forced listeners to select the speech sample that sounds better quality.
      
	\subsection{Experimental results}
        \drawfig{t}{0.98\linewidth}{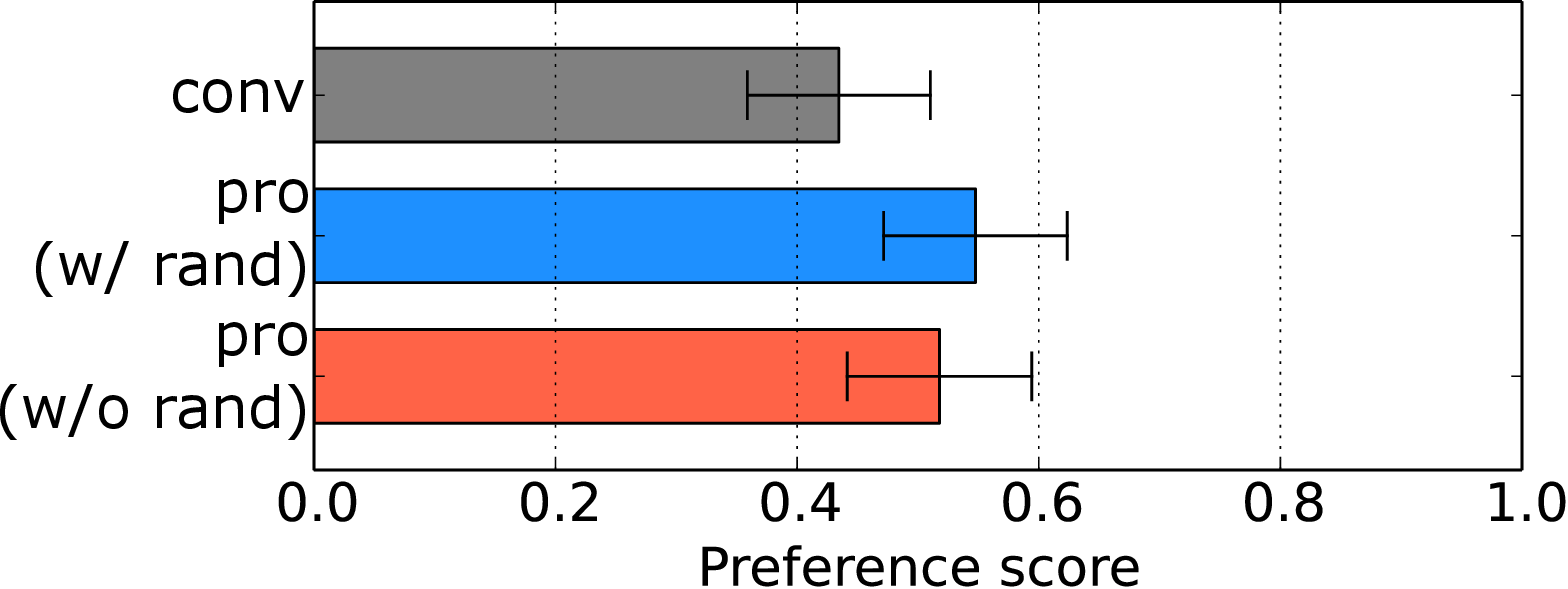}{Preference scores on speech quality with 95\% confidence interval.}

		\Fig{eps/subj} shows the results of the preference test on speech quality.
        We can see that sampling-based generation does not degrade speech quality below that of ML-based generation.
        Also, speech quality of our systems with non-parametric modeling is better than that of conventional speech synthesis assuming an isotropic Gaussian distribution.
        These results demonstrate that the proposed sampling-based generation can generate speech that has inter-utterance variation
        while preserving speech quality.
        Our next work is to compare the variation of natural and synthetic speech, and to apply the algorithm to end-to-end speech synthesis.

\Section{Conclusion}
 This paper presented sampling-based speech parameter generation using moment-matching networks to give synthetic speech natural inter-utterance variation. Although humans' speech has variation within the same linguistic and para-linguistic information, conventional speech synthesis always produces completely the same speech within the same context. To address this gap, we have proposed sampling-based generation using moment-matching networks. The neural networks were trained to transform the simple prior noise to the distribution (variation) of speech parameters. Since the training criterion (conditional maximum mean discrepancy) is non-parametric, our algorithm can model more complicated distributions than conventional mean squared error training assuming an isotropic Gaussian distribution. Furthermore, since the variation is compressed into the simple prior noise, the proposed sampling-based generation is more computationally efficient than direct sampling from distributions of speech parameters. We have investigated whether or not the proposed generation degrades synthetic speech quality. The results demonstrated that our sampling-based generation causes no degradation compared with maximum likelihood-based generation. 
 
	{\bf Acknowledgements:} 
      Part of this work was supported by JSPS KAKENHI Grant Number $16H06681$.

   \eightpt
   \bibliographystyle{IEEEtran}
   \bibliography{tts}

\end{document}